\documentclass[aps,pre,showpacs,superscriptaddress]{revtex4}
\usepackage{graphicx}
\usepackage{natbib}

\def\Mn{\hbox{Mn$_{12}$}}
\def\V{\hbox{V$_{15}$}}

\def\figsize{8.cm}

\begin{document}
\title{Magnetic energy-level diagrams of high-spin (Mn$_{12}$-acetate)\\
and low-spin (V$_{15}$) molecules}
\author{H. De Raedt}
 \email{deraedt@phys.rug.nl}
 \homepage{http://www.compphys.org}
\affiliation{Department of Applied Physics-Computational Physics,
Materials Science Centre, University of Groningen, Nijenborgh 4,
NL-9747 AG Groningen, The Netherlands}
\author{S. Miyashita}
 \email{miya@spin.t.u-tokyo.ac.jp}
\affiliation{Department of Applied Physics, Graduate School of Science,
University of Tokyo, Bunkyo-ku Tokyo 113-8656, Japan}
\author{K. Michielsen}
 \email{kristel@phys.rug.nl}
\affiliation{Department of Applied Physics-Computational Physics,
Materials Science Centre, University of Groningen, Nijenborgh 4,
NL-9747 AG Groningen, The Netherlands}
\begin{abstract}

The magnetic energy-level diagrams for models of the \Mn\ and \V\ molecule
are calculated using the Lanczos method with full orthogonalization
and a Chebyshev-polynomial-based projector method.
The effect of the Dzyaloshinskii-Moriya interaction on the appearance of energy-level repulsions
and its relevance to the observation of steps in the time-dependent
magnetization data is studied.
We assess the usefulness of simplified models for the description of
the zero-temperature magnetization dynamics.
\end{abstract}
\date{\today}

\maketitle

\section{Introduction}\label{sec1}

Magnetic molecules such as \Mn\ or \V\ have attracted a lot of interest recently
because these nanomagnets can be used to study e.g. quantum (de)coherence,
relaxation and tunneling of the magnetization on a nanoscale~\cite{%
Gunter,%
Caneschi,%
gat,%
gat2,%
Levine,%
Friedman,%
Thomas,%
sangregorio,%
fe8tun,%
Bernard,%
Perenboom,%
Irinel3,%
Pohjola,%
Zhong,%
Irinel1,%
Irinel5,%
Irinel2,%
Bouk0,%
Bouk1,%
Wernsdorfer,%
Honecker,%
Irinel4%
}.
As a result of the very weak intramolecular interactions between these molecules,
experiments directly probe the magnetization dynamics of the individual molecules.
In particular the adiabatic change of the magnetization at low-temperature
is governed by the discrete energy-level structure~\cite{Seiji0,Slava0,Gunter0,Hans2}.

The magnetic properties of molecules such as \Mn\ or \V\ are often studied by considering
a simplified model for the magnetic energy levels for a specific spin multiplet,
e.g. S=10 for \Mn\ or S=3/2 for \V.
However for these and other, similar, magnetic molecules that consist of several magnetic moments
(12 in the case of \Mn, 15 in the case of \V),
the reduction of the many-body Hamiltonian to an effective Hamiltonian for a specific spin multiplet is,
except for the diagonal terms, non-trivial.

Magnetic anisotropy, a result of the geometrical arrangement of the magnetic ions within a molecule
of low symmetry, mixes states of different total spin and enforces a treatment of the full Hilbert space of the system.
The dominant contribution to the magnetic anisotropy due to spin-orbit interactions is given by
the Dzyaloshinskii-Moriya interaction
(DMI)~\cite{Dzy,Mor,Kaplan,Shekntman1,Shekntman2,Yosida,Crep}.
In principle this interaction can change energy-level crossings into energy-level repulsions.
The presence of the latter is essential to explain
the adiabatic changes of the magnetization at the resonant fields
in terms of the Landau-Zener-St\"uckelberg (LZS) transition~\cite{Seiji0,Slava0,Gunter0,Hans2}.
Thus a minimal magnetic model Hamiltonian should contain (strong) Heisenberg interactions,
anisotropic interactions and a coupling to the applied magnetic field~\cite{Bernard,
Misha1,%
mn12spl,%
RudraMn12,%
Rudra,%
Seiji1,%
Raghu,%
Hans1,%
Konst,%
RudraSeiji}.

In this paper we calculate the magnetic energy-level diagrams for models of the \Mn\ and \V\ molecule
using exact diagonalization techniques.
We study the effect of the DMI on the appearance of energy-level repulsions that determine
the adiabatic changes of the magnetization observed experimentally.
%MIYA  Rudra also used a full diagonalization partly.
%HANS OK
In contrast to earlier work~\cite{Rudra,Konst},
the approach adopted in the present paper does not rely on perturbation theory.
Instead we perform an exact numerical diagonalization of the full Hamiltonian.

As the quantum spin dynamics of these magnetic molecules is determined by the (tiny) level repulsions,
a detailed knowledge of the low-lying energy levels scheme is necessary.
In order to bridge the energy scales involved (e.g. from 500K, a typical energy scale for the
interaction between individual magnetic ions, to $\approx 10^{-2} - 10^{-9}$K, a typical energy scale for
energy-level splittings),
a calculation of the energy levels of these many-spin Hamiltonians has to be very accurate.
We have tested many different standard algorithms to compute the low-lying states.
For systems that are too large to be solved by full exact diagonalization,
we find that the Lanczos method with full orthogonalization and
a Chebyshev-polynomial-based projector method can solve
these rather large and difficult eigenvalue problems with sufficient accuracy.

%We explore the possibility to derive effective single-spin Hamiltonians based on the knowledge of a number
%(e.g. 21 in the case of \Mn) of the lowest eigenstates of microscopic many-spin models of these molecules.
%We also consider alternative methods to study the quantum dynamics without mapping onto an explicit model
%of the spin multiplet, i.e. by using the low-lying states directly.

The paper is organized as follows.
In Sec.~\ref{sec2} we introduce the model Hamiltonians for the \Mn\ and \V\ molecules.
In Sec.~\ref{sec3} we briefly discuss the numerical algorithms that we use to compute the energy levels.
Our results for the energy level schemes for \Mn\ and \V\ are presented in Sec.~\ref{sec4}.
In Sec.~\ref{sec5} we analyze a reduced, 3-spin model for \V\ and determine the conditions
on the DMI energy-level under which repulsions appear.
Numerical calculations for the full \V\ model confirm
that these conditions are also relevant for the presence of energy-level repulsions in the \V\ model.
%Finally, in Sec.~\ref{sec5} we give our conclusions.

\begin{figure}[t]
\begin{center}
\includegraphics[width=\figsize]{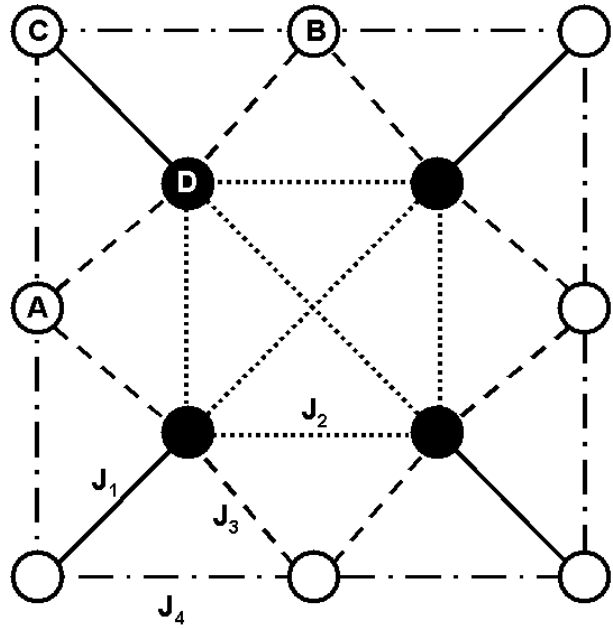}
\caption{%
Schematic diagram of the dominant magnetic (Heisenberg) interactions of the \Mn\ molecule.}
\label{fig1}
\end{center}
\end{figure}

\begin{figure}[t]
\begin{center}
\includegraphics[width=\figsize]{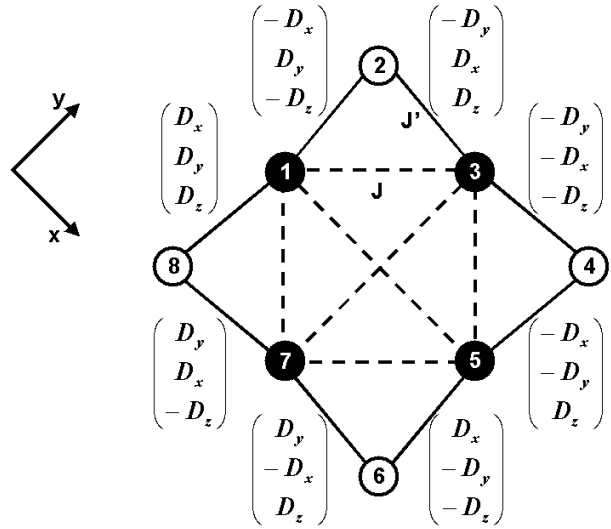}
\caption{%
Schematic diagram of the magnetic interactions of the simplified
model (\ref{MnHam}) of the \Mn\ molecule.}
\label{fig2}
\end{center}
\end{figure}

\section{Models}\label{sec2}
\subsection{Manganese complex: \Mn}

In Fig.~\ref{fig1} we reproduce the
schematic diagram of the dominant magnetic (Heisenberg) interactions of the \Mn\ molecule
(Mn$_{12}$(CH$_{3}$COO)$_{16}$(H$_{2}$O)$_{4}$O$_{12}\cdot$2CH$_{3}$COOH$\cdot$4H$_{2}$O).
The four inner Mn$^{+4}$ ions have a spin $S=3/2$, the other eight Mn$^{+3}$ ions have spin $S=2$.
The number of different spin states of this system is $4^4\times5^8=10^{8}$.
%MIYA
%HANS
If the total magnetization is a conserved quantity, it can be used to
block-diagonalize the Hamiltonian, allowing the study of models of this size~\cite{Raghu,Regnault}.
However, to study the adiabatic change of magnetization, we have to treat
all the states, and the dimension of the matrix become prohibitively large.
Thus we need to simplify the model in order to reduce the dimension.
A drastic reduction of the number of spin states can be achieved by assuming that the strong
antiferromagnetic Heisenberg interaction ($J_1$) between an inner ion and its outer neighbor
allows the replacement of the magnetic moment of an inner ion by an effective S=1/2 moment.
The schematic diagram of this simplified (but still complicated) model is shown in Fig.~\ref{fig2}.
The number of different spin states of this model is $2^4\times5^4=10^{4}$.
In this paper we study the latter model.

The Hamiltonian for the magnetic interactions of the simplified \Mn\ model can be written as~\cite{Misha1}

\begin{eqnarray}
{\cal H}&=& -J\Bigl(\sum_{i=1}^4 {\bf S}_{2i-1} \Bigr)^2
  -J'\sum_{\langle i,j\rangle} {\bf S}_{2i-1} \cdot{\bf S}_{2j}
  -K_z\sum_{i=1}^4 \left(S_{2i}^z\right)^2 %\\ \nonumber  &&
  + \sum_{\langle i,j\rangle} {\bf D}^{i,j} \cdot
     [{\bf S}_{2i-1}\times {\bf S}_{2j}]
     - \sum_{i=1}^{8} {\bf h}\cdot{\bf S}_{i},
\label{MnHam}
\end{eqnarray}
where even (odd) numbered ${\bf S}_i$ are the spin operators for the
outer (inner) $S=2$ ($S=1/2$) spins.
The first two terms describe the isotropic Heisenberg exchange between the spins.
The third term describes the single-ion easy-axis anisotropy of $S=2$ spins.
%MIYA
%HANS
In this paper we do not consider
higher-order correction terms that restore the SU(2) symmetry~\cite{Kaplan,Shekntman1,Shekntman2,Zheludev}.
The fourth term represents the antisymmetric DMI in \Mn.
The vector ${\bf D}^{i,j}$ determines the DMI between the $i$-th $S=1/2$ spin and the $j$-th $S=2$ spin.
The last term describes the interaction of the spins with the external field ${\bf h}$.
Note that the factor $g\mu_B$ is absorbed in our definition of ${\bf h}$.

The first three terms in Hamiltonian (\ref{MnHam}) conserve the
$z$-component of the total spin $M_z=\sum_{i=1}^{8} S^z_{i}$.
The DMI on the other hand mixes states with different total spin and also states with the same total spin.
Hence, the DMI can change level crossings into level repulsions.
Therefore, the presence of the DMI may be sufficient
to explain the experimentally observed adiabatic changes of the magnetization.

The four-fold rotational-reflection symmetry ($S_4$)
of the \Mn\ molecule imposes some relations between the DM-vectors.
It follows that there are only three independent DM-parameters:
$D_x \equiv D_x^{1,8}$, $D_y\equiv D_y^{1,8}$, and $D_z \equiv D_z^{1, 8}$,
as indicated in Fig.~\ref{fig2}.
The above model satisfactorily describes a rather wide range of experimental data,
such as the splitting of the neutron scattering peaks, results of EPR measurements and the
temperature dependence of magnetic susceptibility~\cite{Misha1}.
The parameters of this model have been estimated by comparing experimental and theoretical data.
In this paper we will use the parameter set B from Ref.~\cite{Misha1,Hans1}:
$J=23.8$K, $J'=79.2$K, $K_z=5.72$K, $D_x=22$K, $D_y=0$, and $D_z=10$K.

Although the amount of available data is not sufficient to fix all these parameters accurately,
we expect that the general trends in the energy-level diagram will not change drastically
if these parameters change relatively little.

\begin{figure}[t]
\begin{center}
\includegraphics[width=6cm]{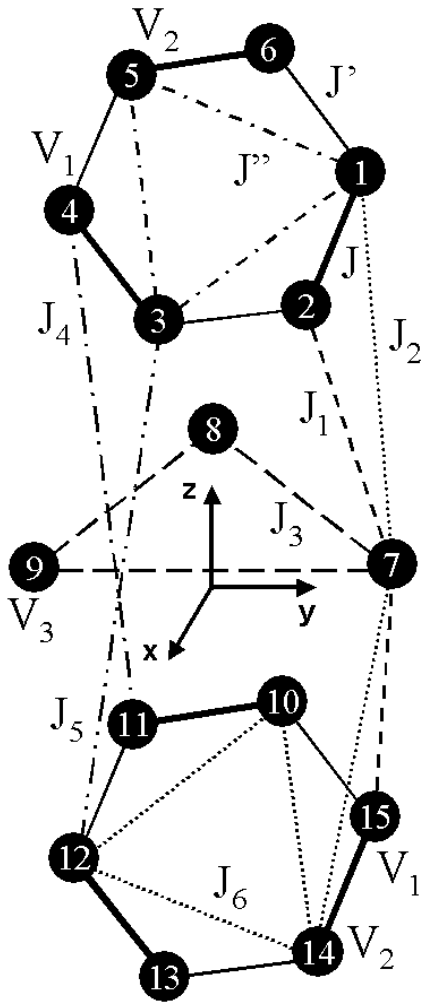}
\caption{%
Schematic diagram of the magnetic interactions in model (\ref{VHam}) of the \V\ molecule.}
\label{fig3}
\end{center}
\end{figure}

\subsection{Vanadium complex: \V}
In Fig.~\ref{fig3} we show the schematic diagram of the dominant magnetic (Heisenberg) interactions of
the \V\ molecule
(K$_{6}$[V$^{IV}_{15}$As$_{6}$O$_{42}$(H$_{2}$O)]$\cdot$8H$_{2}$O).
The magnetic structure consists of two hexagons with six S=1/2 spins each, enclosing a triangle with
three S=1/2 spins. All dominant Heisenberg interactions are antiferromagnetic.
The number of different spin states of this model is $2^{15}=32768$.
%MIYA %HANS: OK
The minimal Hamiltonian for the magnetic interactions
that incorporates the effects on magnetic anisotropy can be written as~\cite{Seiji1,Rudra,Konst,Irinel4}

\begin{eqnarray}
{\cal H}&=& -\sum_{\langle i,j\rangle} J_{i,j}{\bf S}_k \cdot{\bf S}_l
  + \sum_{\langle i,j\rangle} {\bf D}^{i,j}\cdot[{\bf S}_i\times {\bf S}_j]
     - \sum_{i} {\bf h}\cdot S^z_{i}.
\label{VHam}
\end{eqnarray}

The various Heisenberg interactions $J_{i,j}$ are shown in Fig.~\ref{fig3}.
For simplicity, we assume that ${\bf D}^{i,j}=0$ for sites $i$ and $j$
except for bonds for which the Heisenberg exchange constant is $J$ %MIYA %HANS OK
(see Fig.~\ref{fig3})~\cite{Konst,Rudra}.
Rotations about $2\pi/3$ and $4\pi/3$ around the axis perpendicular
to and passing through the center of the
hexagons leave the \V\ complex invariant.
This enforces constraints on the values of ${\bf D}^{i,j}$~\cite{Konst,RudraSeiji}.
In Sec.~\ref{sec4} we present results for
several different sets of estimates for the model parameters
of the \V\ model~\cite{gat2,Rudra,Konst,Bouk0}.

\section{Numerical method}\label{sec3}

A theoretical description of quantum dynamical phenomena in the \Mn\ and \V\ nanomagnets
requires detailed knowledge of their energy-level schemes.
Disregarding the fascinating physics of the nanomagnets, the calculation of the eigenvalues
of their model Hamiltonians is a challenging problem in its own right.
Firstly, the (adiabatic) quantum dynamics of these systems is mainly determined by the (tiny) level repulsions.
Therefore the calculation of the energy levels of these many-spin Hamiltonians has to be very accurate
in order to bridge the energy scales involved (e.g. from 500K to $\approx 10^{-9}$K).
Secondly, the level repulsions originate from the DMI that mix states with different magnetization.
In principle, this prevents the use of the magnetization as a vehicle to block-diagonalize the Hamiltonian
and effectively reduce the size of the matrices that have to be diagonalized.
If a level repulsion involves states of significantly different magnetization (e.g. $M^z=-10$ and $M^z=10$)
a perturbative calculation of the level splitting would require going to rather high order (at least 20),
a cumbersome procedure. Therefore it is of interest to explore alternative routes to
direct but accurate, brute-force diagonalization of the full model Hamiltonian.

As a non-trivial set of reference data, we used the eigenvalues obtained
by full diagonalization (using standard LAPACK algorithms)
of the $10000\times10000$ matrix representing model (\ref{MnHam}) \cite{Hans1}.
For one set of model parameters, such a calculation takes about 2 hours of CPU time on an
Athlon 1.8 GHz/1.5Gb system. Clearly this is too slow if we want to compute the energy-level diagram
%MIYA %HANS, yes it sounds better to us too
%for many different ${\bf h}$-fields.
as a function of the magnetic field ${\bf h}$.
% For me ${\bf h}$-fields sounds strange. Can we use simply ${\bf h}$?
%
In particular if we want to estimate the structure of the level splittings
at the resonant fields we need the eigenvalues 
%MIYA
%for many ${\bf h}$-fields.
for many values of ${\bf h}$.
Furthermore, in the case of \V\ this calculation would take
about 30 times longer and require about 15 Gb of memory which, for present-day computers,
is too much to be of practical use.

We have tested different standard algorithms to compute the low-lying eigenvalues of large matrices.
The standard Lanczos method (including its conjugate gradient version)
as well as the power method~\cite{WILKINSON,GOLUB}
either converge too slowly, lack the accuracy to resolve the (nearly)-degenerate eigenvalues,
and sometimes even completely fail to correctly reproduce the low-lying part of the spectrum.
This is not a surprise: by construction these methods work well if the ground state is not degenerate
and there is little guarantuee that they will work if there are (nearly)-degenerate
eigenvalues~\cite{WILKINSON,GOLUB}.
In particular, the Lanczos procedure suffers from numerical instabilities due to the loss of
orthogonalization of the Lanczos vectors~\cite{WILKINSON,GOLUB}.
It seems that model Hamiltonians for the nanoscale magnets provide a class of (complex Hermitian)
eigenvalue problems that are hard to solve.

Extensive tests lead us to the conclusion that only the Lanczos method with full orthogonalization
(LFO)~\cite{WILKINSON,GOLUB}
and a Chebyshev-polynomial-based projector method (CP) (see Appendix) can solve
these rather large and difficult eigenvalue problems with sufficient accuracy.
The former is significantly faster than the latter but using both gives extra confidence in the results.

In the Lanczos method with full orthogonalization, each time a new Lanczos vector is generated
we explicitly orthogonalize (to working precision) this vector to all, not just to the two previous,
Lanczos vectors~\cite{WILKINSON,GOLUB}.
With some minor modifications to restart the procedure when the Lanczos iteration terminate
prematurely, after $n$ steps this procedure tranforms $n\times n$ matrix $H$
into a tri-diagonal matrix that is comparable in accuracy
to the one obtained through Householder tri-diagonalization but offers no advantages~\cite{GOLUB}.
In our case we are only interested in a few low-lying eigenstates of $H$.
Thus we can exploit the fact that projection onto the (numerically exact) subspace of dimension $k$ ($k \ll n$),
built by the Lanczos vectors will yield increasingly accurate estimates
of the smallest (largest) eigenvalues and corresponding eigenvectors as $k$ increases.

In practice, to compute the $M$ lowest energy levels, the LFO procedure is carried out as follows.
\begin{itemize}
\item Perform a Lanczos step according to the standard procedure
\item Use the modified Gramm-Schmidt procedure to orthogonalize the new Lanczos vector with respect to all
previous ones~\cite{WILKINSON,GOLUB}
\item Compute the matrix elements of the tridiagonal matrix
\item At regular intervals, diagonalize the tridiagonal matrix, compute the
approximate eigenvectors $\varphi_i$,
$\mu_i=\langle\varphi_i|H|\varphi_i\rangle$ and
$\Delta^2_i=\langle\varphi_i|(H-\mu_i)^2|\varphi_i\rangle$ for $i=1,\ldots,M$, and check if all
$\Delta_i$ are smaller that a specified threshold. If so, terminate the procedure (the exact
eigenvalue $E_i$ closest to $\mu_i$ satisfies $\mu_i-\Delta \le E_i\le \mu_i+\Delta_i$). If not, continue generating new
Lanczos vectors, etc.
\end{itemize}
%MIYA %HANS: see table caption, full diagonalization is mentioned above.
% question  How fast the new methods compared with full-diagionalization?
%

\section{Results}\label{sec4}

\subsection{Manganese complex: \Mn}
\begin{figure}[t]
\begin{center}
\includegraphics[width=\figsize]{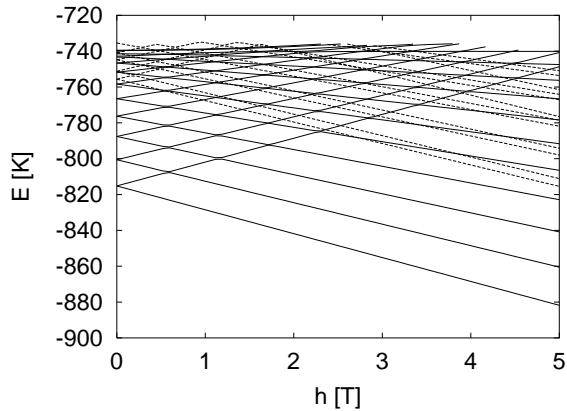}
\caption{%
The lowest 21 energy levels of the \Mn\ model (\ref{MnHam}) as a function of the applied magnetic field
${\bf h}$.
Solid lines: eigenstates with $|M^z|\approx10$;
dashed lines: eigenstates with $|M^z|\approx9$.
}
\label{fig4}
\end{center}
\end{figure}
\begin{table}[t]
\begin{center}
\caption{%
The 21 lowest eigenvalues $E_i$ and total spin $S_i$ of the corresponding eigenstates of
the \Mn\ model (\ref{MnHam}) for two values of the external applied field ${\bf h}$ along the $z$-axis.
The distance between $E_i$ and the exact eigenvalue closest to $E_i$ is
$\Delta_i=\langle\varphi_i|(H-E_i)^2|\varphi_i\rangle^{1/2}<10^{-10}$ for $i=1,\ldots,7$.
These calculations took about 20 minutes on an Athlon 1.8 GHz/1.5Gb system, using
1000 fully orthogonal Lanczos vectors.}
\label{Mntab}
\begin{ruledtabular}\begin{tabular}{ccccc}
$i$ & $E_i(h=0)$ & $S_i(h=0)$ & $E_i(h=5T)$ & $S_i(h=5T)$ \\
\hline
  0 &  -815.1971469173  &    9.91 & -881.7827744750   & 9.92 \\
  1 &  -815.1971469173  &    9.91 & -860.4928253394   & 9.93 \\
  2 &  -800.5810020061  &    9.91 & -840.8569089483   & 9.92 \\
  3 &  -800.5810020061  &    9.91 & -822.8556918884   & 9.92 \\
  4 &  -787.6124037484  &    9.91 & -815.4339009404   & 8.93 \\
  5 &  -787.6124037482  &    9.91 & -811.0766283789   & 8.93 \\
  6 &  -776.2715579413  &    9.90 & -806.4609011890   & 9.90 \\
  7 &  -776.2715579281  &    9.90 & -797.9409264313   & 8.94 \\
  8 &  -766.5314713958  &    9.90 & -794.1268159385   & 8.93 \\
  9 &  -766.5314702412  &    9.90 & -791.6387794071   & 9.90 \\
 10 &  -758.3618785887  &    9.89 & -781.8373616760   & 8.93 \\
 11 &  -758.3618126323  &    9.89 & -778.4824830935   & 8.96 \\
 12 &  -755.6412882369  &    8.92 & -778.3500886860   & 9.85 \\
 13 &  -755.6412882368  &    8.92 & -776.4751565103   & 8.93 \\
 14 &  -751.7362729420  &    9.88 & -767.0677893890   & 8.93 \\
 15 &  -751.7337526641  &    9.88 & -766.5785427469   & 9.87 \\
 16 &  -751.2349837637  &    8.91 & -764.0838038821   & 8.92 \\
 17 &  -751.2349837632  &    8.91 & -761.4314952668   & 8.76 \\
 18 &  -746.6655233754  &    9.87 & -756.2910279030   & 9.87 \\
 19 &  -746.6082906321  &    9.87 & -753.5740765004   & 8.92 \\
 20 &  -744.8208087762  &    8.92 & -752.7461619357   & 8.08 \\
 \end{tabular}
 \end{ruledtabular}
 \end{center}
 \end{table}

In Table~\ref{Mntab} we present the numerical data for $h=0$T and $h=5$T, also obtained
by LFO. The results obtained by full exact diagonalization (LAPACK), LFO and CP are
the same to working precision (about 13 digits).
In Fig.~\ref{fig4} we show the results for the lowest 21 energy levels of the
\Mn\ model as a function of the applied magnetic field as obtained by LFO.
%MIYA %HANS: we changed the order of the sentences, we think it is less ambigious now.
%In this way one can calculate the energy levels with the full-diagonalization
%method. However it takes very long time and it is not adequate to study
%the field dependence as we mentioned before. Thus here we use the
%Lanczos method with orthogonaization.
% question: the data for Fig. 4 is obtained by full-diagonalization?
%           or Loncos+orthogonalization?
%HANS: LFO, now mentioned in first sentence of paragraph

Although the total magnetization is not a good quantum number, we can
label the various eigenstates by their (calculated) magnetization.
For large fields and/or energies, eigenstates with total spin 8, 9 and 10 appear,
as shown in Table~\ref{Mntab}.
In Fig.~\ref{fig4} eigenstates with $|M^z|\approx10 (9)$ (within an error of about 10\%)
are represented by solid (dashed) lines (eigenstates with $|M^z|\approx8$ appear for $h>4$ but
have been omitted for clarity).

The standard $S=10$ single-spin model for \Mn\
\begin{eqnarray}
{\cal H}&=& -D (S^z)^2-h S^z,
\label{SSHam}
\end{eqnarray}
is often used as a starting point to interpret experimental
results~\cite{Friedman,Thomas,Perenboom,Irinel3,Pohjola,Rudra}.
The energy levels of this model exhibit crossings at the resonant fields $h=\pm Dn$ for $n=-10,\ldots,10$,
in agreement with our numerical results for the more microscopic model (\ref{MnHam}).
For the parameter set B, we find that $D\approx0.55K$, in good agreement with
experiments~\cite{Friedman,Thomas}.

The single-spin model (\ref{SSHam}) commutes with the magnetization $S^z$ and
therefore it only displays level crossings, no level repulsions.
Adding an anisotropy term of the form $S_+^4 + S_-^4$ only
leads to level repulsions when the magnetization changes by 4, which does not agree with the observation
of adiabatic changes of the magnetization for all $h=nD$~\cite{Friedman,Thomas,Perenboom,Irinel3}.
In contrast, for the DMI the Hamiltonian has nonzero matrix elements for the pairs of states
$|S,S_z\rangle$ and  $|S\pm 1,S_z\pm1\rangle$, but zero matrix elements for levels
with the same value of the total spin.

In Fig.~\ref{fig4}, for some values of $h$, level repulsions are present.
However, these are due to the fitting procedure used to plot the data
and the number of $h$-values used (100) and
disappear by using a higher resolution in ${\bf h}$-fields (results not shown).
Thus these splittings have no physical meaning.
For the \Mn\ system, the energy splittings at low field are extremely small.
Their calculation requires extended-precision (128-bit) arithmetic \cite{Hans1}.
Therefore, to study the structure of the energy-level diagram in more detail
we concentrate on the transitions at $h\approx3.4$T
($M_z\approx-10\rightarrow M_z\approx4$)and $h\approx3.9$T
($M_z\approx-10\rightarrow M_z\approx3$)
for which adiabatic changes of the magnetization have been observed
in experiments~\cite{Friedman,Thomas,Perenboom,Irinel3}.
From experiments one finds that the magnitude of
these splittings is of the order of 10 nK~\cite{private}.
Extensive calculations lead us to the conclusion that the energy splitting at
these resonant fields is smaller than $10^{-6}$K.
%MIYA
% question
% nK is 10^{-9}K which is much smaller than 10^{-6}K. Thus your observation
% below seems natural??
%HANS: yes, we think so. It is not a surprise.
Adding an extra transverse field by tilting the ${\bf h}$-field by 5 degrees
does not change this conclusion.
Thus, it is clear that within the (very high) resolution in the ${\bf h}$-field and
13-digit precision of the calculation, there is no compelling evidence that the DMI
gives rise to a level repulsion, at least not
for the choice of model parameters (set B, see above) considered here.
The algorithms developed for the work presented in this paper can
be used for 33-digit calculations without modification
and we leave the calculation of the splittings for future work.

%As this observation on the basis of our numerical results is more generic than
%it may seem at this point, we postpone a more detailed analysis until we have
%examined the energy-level structure of the \V\ model.

\subsection{Vanadium complex: \V}

\begin{figure}[t]
\begin{center}
\setlength{\unitlength}{1cm}
\begin{picture}(14,6)
\put(-1.5,0.){\includegraphics[width=8cm]{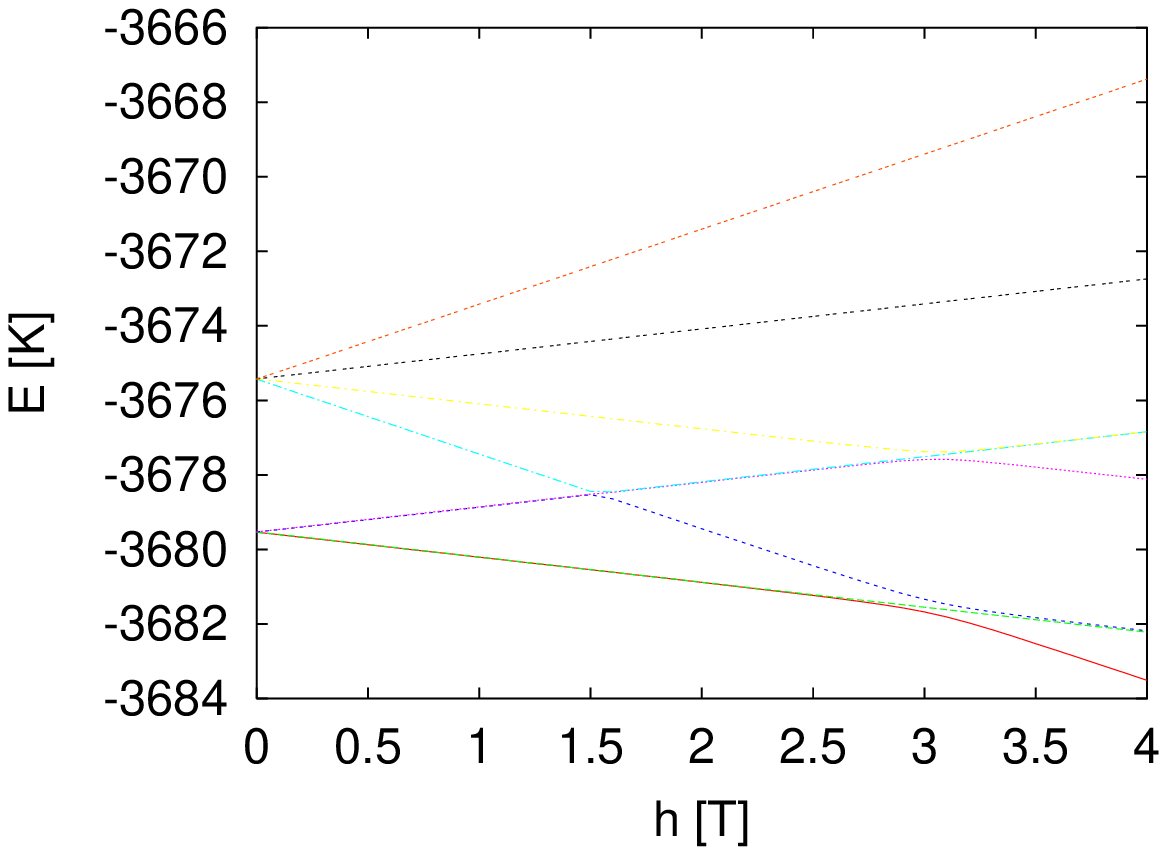}}
\put(7.,0.){\includegraphics[width=8cm]{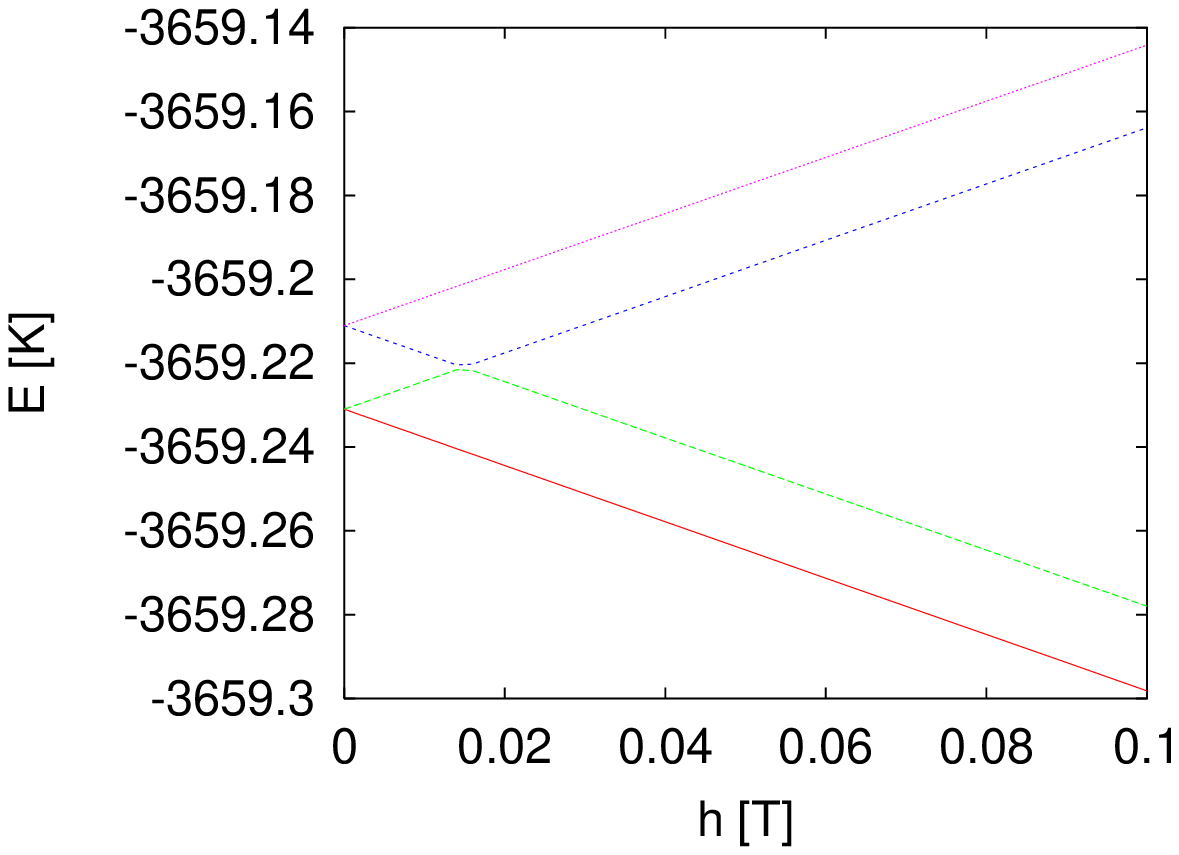}}
\end{picture}
\caption{%
Left:
The lowest 8 energy levels of \V\ model (\ref{VHam})
with model parameters taken from Ref.~\cite{Rudra} ({\bf VsetA})
as a function of the applied magnetic field ${\bf h}$
parallel to the $z$-axis.
Right: Detailed view of the four lowest energy levels at $h\approx0$.}
\label{fig5}
\end{center}
\end{figure}

\begin{table}[t]
\begin{center}
\caption{%
The eight lowest eigenvalues $E_i$ and total spin $S_i$ of the corresponding eigenstates of
the \V\ model (\ref{VHam})
with model parameters taken from Ref.~\cite{Rudra} ({\bf VsetA})
for two values of the external applied field ${\bf h}$ parallel the $z$-axis.
The distance between $E_i$ and the exact eigenvalue closest to $E_i$ is
$\Delta_i=\langle\varphi_i|(H-E_i)^2|\varphi_i\rangle^{1/2}<\times10^{-9}$ for $i=1,\ldots,7$.
These calculations took less than 20 minutes on a Cray SV1 computer, using
521 fully orthogonal Lanczos vectors.}
\label{Vtab}
\begin{ruledtabular}\begin{tabular}{ccccc}
$i$ & $E_i(h=0)$ & $S_i(h=0)$ & $E_i(h=4T)$ & $S_i(h=4T)$ \\
\hline
  0 & -3679.53623744 &  0.51 &-3683.51181131& 1.50 \\
  1 & -3679.53623744 &  0.51 &-3682.21997451& 0.51 \\
  2 & -3679.52777009 &  0.51 &-3682.18488706& 0.53 \\
  3 & -3679.52777009 &  0.51 &-3678.11784886& 1.50 \\
  4 & -3675.42943612 &  1.50 &-3676.84225573& 0.52 \\
  5 & -3675.42943612 &  1.50 &-3676.83951808& 0.51 \\
  6 & -3675.42325141 &  1.50 &-3672.74011178& 1.50 \\
  7 & -3675.42325141 &  1.50 &-3667.37940477& 1.50 \\
 \end{tabular}
 \end{ruledtabular}
 \end{center}
 \end{table}

%As a check on our calculations we performed calculations with
%several sets of model parameters.
%Using the estimates $J=-800$, $J_1=J'=-150$K, and $J_2=J''=-300$K ~\cite{gat2},
%we obtain an energy gap of 5.2K, substantially higher than the experimental value.

For the model parameters given in Ref.~\cite{Rudra},
$J=-800$, $J_1=J'=-54.4$K, and $J_2=J''=-160$K, $J_3=J_4=J_5=J_6=0$ and
in the absence of the DMI,
we find for the energy gap between the ground state
and the first excited state at $h=0$ a value of 4.12478K, in perfect agreement with Ref.~\cite{Rudra}.
Following Ref.~\cite{RudraSeiji} we take for the DMI parameters
$D_x^{1,2}=D_y^{1,2}=D_z^{1,2}=40$K,
which is approximately 5\% of the largest Heisenberg coupling.
Using the rotational symmetry of the hexagon we have
$D_x^{3,4}=14.641$K, $D_y^{3,4}=-54.641$K, $D_z^{3,4}=40$K and
$D_x^{5,6}=-54.641$K, $D_y^{5,6}=14.641$K, $D_z^{5,6}=40$K.
% $D_x^{3,4}=5.8564$K, $D_y^{3,4}=-21.856$K, $D_z^{3,4}=16K and
% $D_x^{5,6}=-21.856$K, $D_y^{5,6}=-5.8564$K, $D_z^{5,6}=16$K.
As the two hexagons are not equivalent we cannot use
symmetry to reduce the number of free parameters.
For simplicity, we assume that the $(x,y)$ positions of the spins on the lower hexagons
differ from those on the upper hexagon by a rotation about $\pi/3$.
This yields for the remaining model parameters
$D_x^{10,11}=-14.641$K, $D_y^{10,11}=54.641$K, $D_z^{10,11}=40$K,
$D_x^{12,13}=-40$K, $D_y^{12,13}=-40$K, $D_z^{12,13}=40$K, and
$D_x^{14,15}=54.641$K, $D_y^{14,15}=-14.641$K, $D_z^{14,15}=40$K.
% $D_x^{10,11}=-5.8564$K, $D_y^{10,11}=21.856$K, $D_z^{10,11}=16$K,
% $D_x^{12,13}=-16$K, $D_y^{12,13}=-16$K, $D_z^{12,13}=16$K, and
% $D_x^{14,15}=21.856$K, $D_y^{14,15}=-5.8564$K, $D_z^{14,15}=16$K.
We will refer to this choice as {\bf VsetA}.
In Fig.~\ref{fig5} we show the results for
the eight lowest energy levels of \V\ model (\ref{VHam})
as a function of the applied magnetic field along the $z$-axis,
using the parameters {\bf VsetA}.

From Table~\ref{Vtab} we see that for zero field, the DMI splits the
doubly-degenerate doublet of $S=1/2$ states into two doublets of $S=1/2$ states.
The difference in energy between the doubly-degenerate, first excited states and the
two-fold degenerate ground states is due to the DMI and,
for the parameters {\bf VsetA}, has a value of $\approx0.0085$K,
much smaller than the experimental estimate $\approx0.05$K~\cite{Irinel4}, but
of the same order of magnitude as the values cited in Ref.~\cite{Konst}.
The next four higher levels are $S=3/2$ states.
The energy-level splitting between the $S=3/2$ and $S=1/2$ states is
$\approx 4.1$K, in reasonable agreement with the experimental value
$\approx 3.7$K~\cite{private}.

\begin{figure}[t]
\begin{center}
\setlength{\unitlength}{1cm}
\begin{picture}(14,6)
\put(-1.5,0.){\includegraphics[width=8cm]{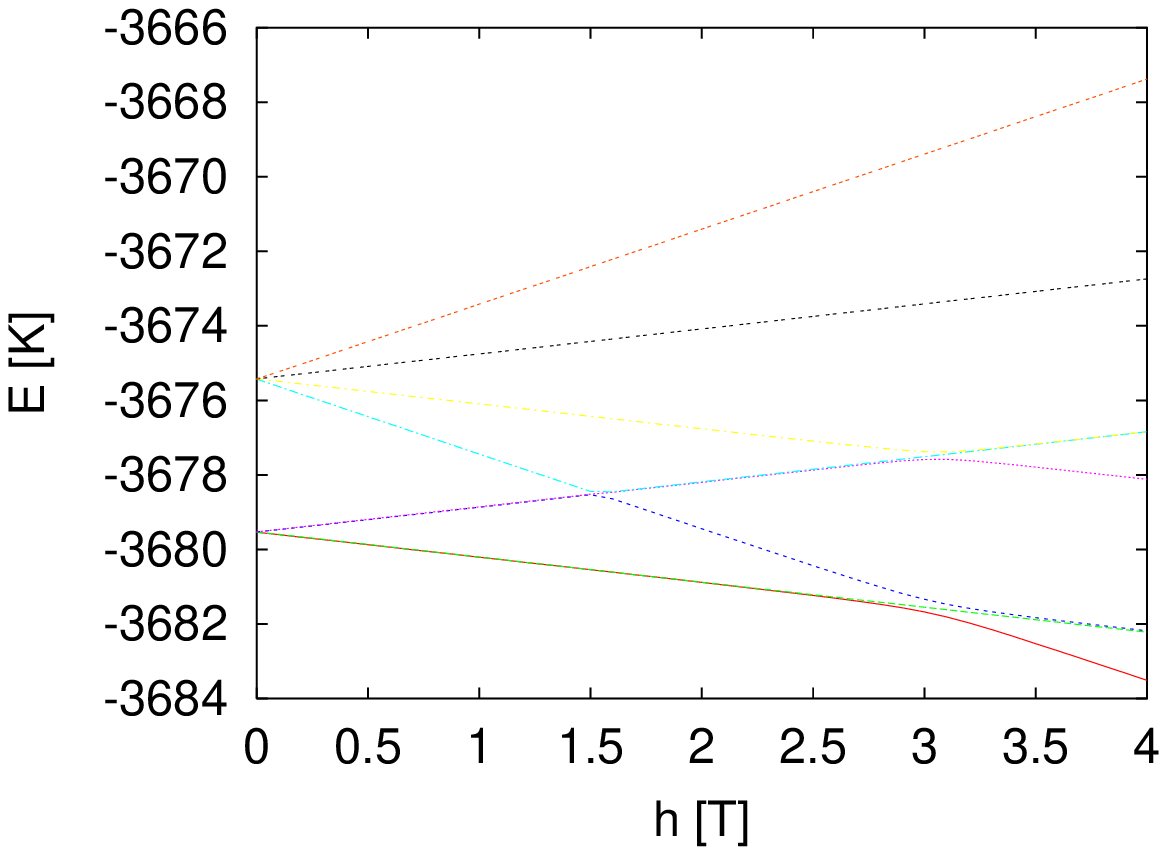}}
\put(7.,0.){\includegraphics[width=8cm]{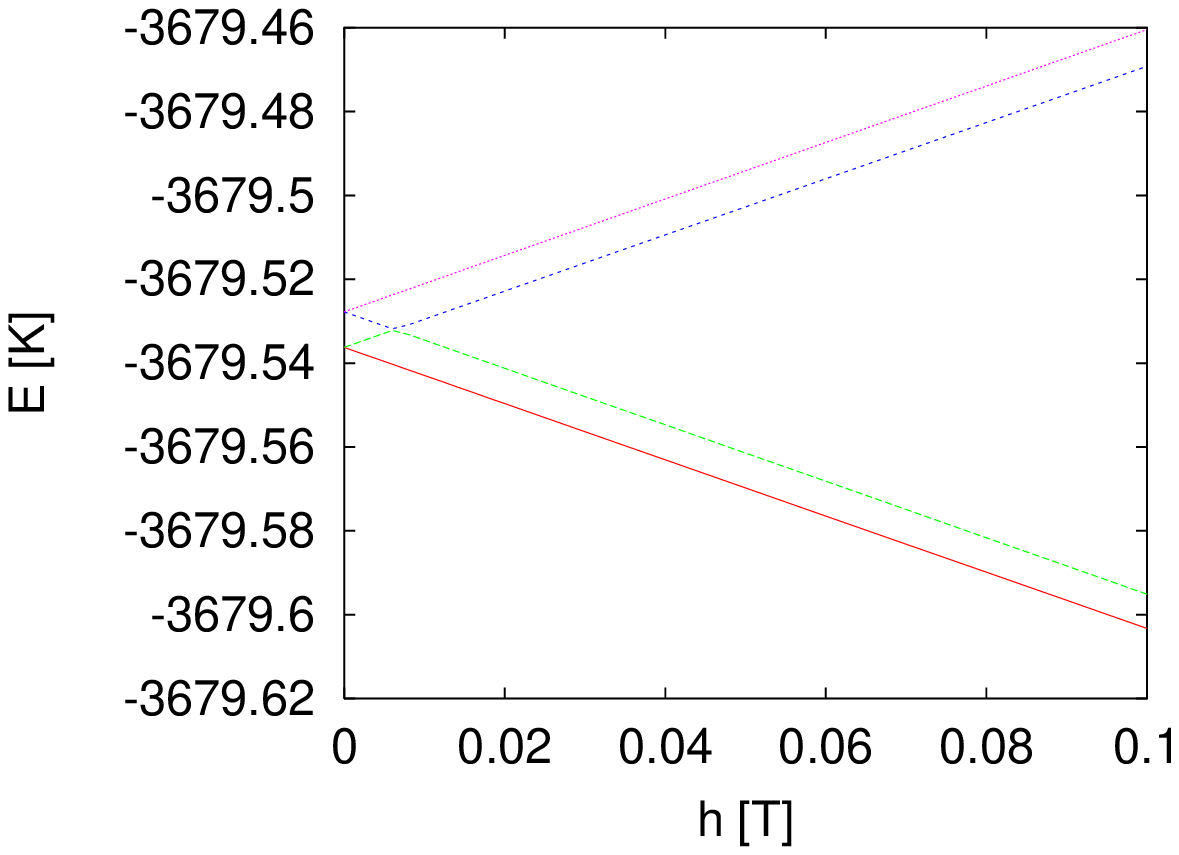}}
\end{picture}
\caption{%
Left:
The lowest 8 energy levels of \V\ model (\ref{VHam})
with model parameters taken from Ref.~\cite{Konst} ({\bf VsetB})
as a function of the applied magnetic field ${\bf h}$
parallel to the $z$-axis.
Right: Detailed view of the four lowest energy levels at $h\approx0$.}
\label{fig6}
\end{center}
\end{figure}

%MIYA
Following Ref.~\cite{Konst}, we take %for the non-zero Heisenberg interactions
$J=-800$, $J_1=J'=-225$K, $J_2=J''=-350$K, and $J_3=J_4=J_5=J_6=0$. %HANS: we should then mention the other J's
In the absence of a DMI, we find that the energy gap between the four-fold degenerate ground state
and the first excited state is 3.61K, in full agreement with the result of Ref.~\cite{Konst}.
Note that this value of the gap is fairly close to the experimental value
of 3.7K~\cite{private}.
Taking for the non-zero DMIs $D_x^{1,2}=D_x^{14,15}=25$K,
$D_x^{3,4}=D_x^{5,6}=D_x^{10,11}=D_x^{12,13}=-12.5$K,
$D_y^{3,4}=-D_y^{5,6}=-D_y^{10,11}=D_y^{12,13}=-21.5$K,
our calculation for the splitting between the two doubly-degenerate
S=1/2 levels yields $0.0037$K, about a factor of two
larger than the value cited in Ref.~\cite{Konst}.
For the energy splitting between the $S=1/2$ and $S=3/2$ levels
we obtain $3.616$K instead of the value $3.618$K given in Ref.~\cite{Konst}.
These differences seem to suggest that a perturbation approach for the DMI
has to be applied with great care~\cite{kosty}.
In Fig.~\ref{fig6} we show the results
for $J=-800$, $J_1=J'=-225$K, and $J_2=J''=-350$K~\cite{Konst} and
the same DMI parameters as in {\bf VsetA} (which we will refer to as {\bf VsetB}).

For the energy gap at zero field,
we find 4.1K and 3.61K
for {\bf VsetA} and {\bf VsetB}
respectively
whereas the experimental estimate is 3.7K~\cite{private}.
The transition between the states $|1/2,1/2\rangle$ and $|3/2,3/2\rangle$ takes place
at $h\approx2.8$T and $h\approx3.0$T respectively, also in good agreement with the experimental
value $2.8$T.

The most advanced estimation of the model parameters {\bf VsetC} is given in Ref.~\cite{Bouk0}.
Taking $J=-809$, $J'=-120$K, $J''=120$K, $J_1=-30$K, $J_2=-122$K,
$J_3=-3$K, $J_4=-11$K, $J_5=-3$K, $J_6=-2K$ (see Table I in Ref.~\cite{Bouk0})
yields an energy gap of 4.915K, in agreement with Ref.~\cite{Bouk0}.
At $h\approx3.6$T, the $S=1/2$ and $S=3/2$ states mix, a level repulsion appears
and the adiabatic change of the magnetization from $M\approx1/2$ to $M\approx3/2$
gives rise to a step in the magnetization versus (time-dependent) ${\bf h}$-field.
Although the qualitative features of the energy-level diagram for {\bf VsetC}
also agree with what one would expect on the basis of experiments,
the field at which the states $|1/2,1/2\rangle$
and $|3/2,3/2\rangle$ cross, $h\approx3.6$T,
does not compare well to the experimental estimate $h\approx2.8$T.

\begin{figure}[t]
\begin{center}
\setlength{\unitlength}{1cm}
\begin{picture}(14,6)
\put(-1.5,0.){\includegraphics[width=8cm]{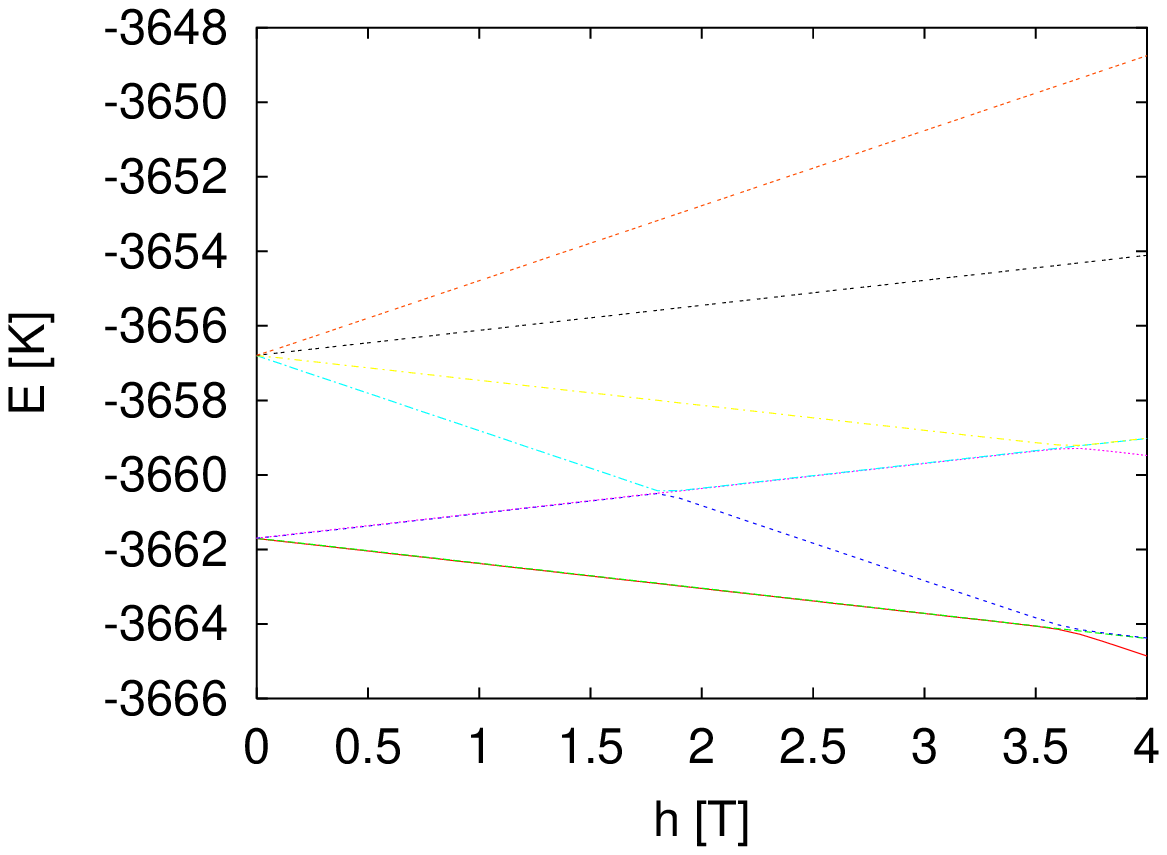}}
\put(7.,0.){\includegraphics[width=8cm]{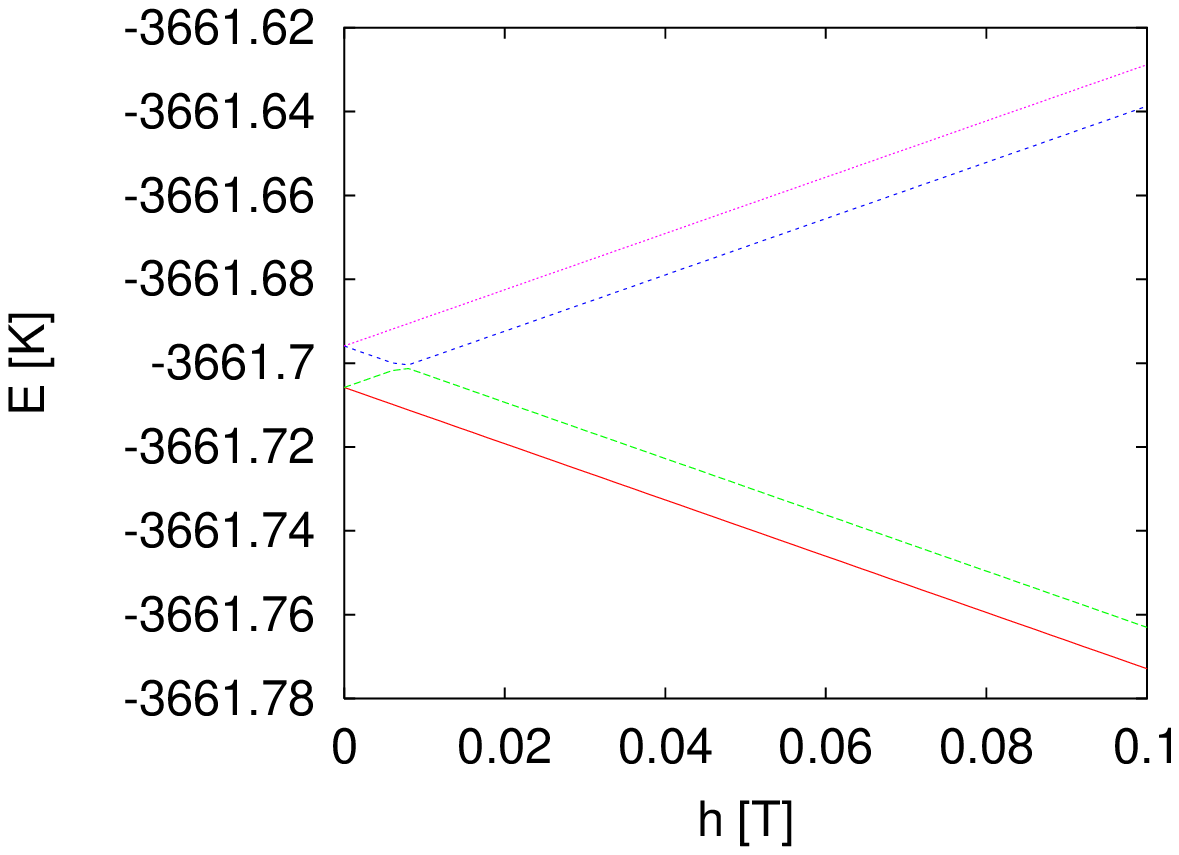}}
\end{picture}
\caption{%
Left:
The lowest 8 energy levels of \V\ model (\ref{VHam})
with model parameters taken from Ref.~\cite{Bouk0} ({\bf VsetC})
as a function of the applied magnetic field $h$
parallel the $z$-axis.
Right: Detailed view of the four lowest energy levels at $h\approx0$.}
\label{fig7}
\end{center}
\end{figure}

On a coarse scale, the level diagrams for {\bf VsetA}, {\bf VsetB} and {\bf VsetC}
are all similar and also resemble those of Ref.~\cite{Konst}.
However, on a finer $h$-scale a new feature appears (see right panel of
Figs.~\ref{fig5},~\ref{fig6}, and \ref{fig7}): the field at which the energy difference
between the second and third level reaches a minimum is no longer at $h=0$.
In other words, in the presence of the DMI, the adiabatic transition between
the states $|1/2,-1/2\rangle$ and 
%MIYA $|1/2,3/2\rangle$ %HANS, yes of course
$|1/2,1/2\rangle$ does not occur. 
%at $h=0$ but at a non-zero value of ${\bf h}$.
%
As we show in the next section, this seems to be a generic feature of the DMI
in models of \V.

%The fact that this effect is not seen in our results for {\bf VsetC} is due
%to our choice of the strength of the DMI.
%Increasing $D_x^{1,2}=D_y^{1,2}=D_z^{1,2}$ from $16$K to 40$K$ yields a diagram
%that is very similar to the left picture in Fig.~\ref{fig8}.

\section{Discussion}\label{sec5}

As shown above, the effect of the DMI on the energy-level diagram is much
larger for the \V\ model than it is for the \Mn\ model.
Therefore, to study the possibility of using simplified models for capturing
the essential time-dependent magnetization dynamics,
we will focus on models for the \V\ molecule which is somewhat easier to treat
numerically.
Qualitatively the energy-level scheme for the eight lowest energy levels
of the \V\ models considered in Sec.~\ref{sec4} closely resembles the
energy-level diagram of a reduced, anisotropic model
of three $S=1/2$ spins described by the Hamiltonian~\cite{Irinel1,Seiji1,Konst}

\begin{eqnarray}
{\cal H}&=& -J\left(
{\bf S}_{1}\cdot{\bf S}_{2}+
{\bf S}_{2}\cdot{\bf S}_{3}+
{\bf S}_{1}\cdot{\bf S}_{3}\right)
%\\ \nonumber  &&
+  {\bf D}^{1,2} \cdot [{\bf S}_{1}\times {\bf S}_{2}]
+  {\bf D}^{2,3} \cdot [{\bf S}_{2}\times {\bf S}_{3}]
+  {\bf D}^{1,3} \cdot [{\bf S}_{1}\times {\bf S}_{3}]
- \sum_{i=1}^{3} {\bf h}\cdot{\bf S}_{i}.
\label{H3}
\end{eqnarray}
%
%We now address the question to what extent model (\ref{H3}) can provide a consistent description
%of the energy-level diagram of the \V\ molecule.

In the absence of the DMI, fitting the energy-level diagram
of model (\ref{H3}) to exprimental data yields $J\approx-2.5$K~\cite{Irinel1}.
We use this estimate to fix $J$ in our numerical calculations.
The number of free parameters can be reduced further by
exploiting the rotational symmetry of the triangle.
We have
$D^{1,2}_x=D_x$, $D^{1,2}_y=D_y$,
$D^{2,3}_y= -(\sqrt{3}D_x+D_y)/2$,
$D^{1,3}_x= -(D_x+\sqrt{3}D_y)/2$,
$D^{1,3}_y= (\sqrt{3}D_x-D_y)/2$, and
$D^{1,2}_z=D^{2,3}_z=D^{1,3}_z=D_z$.
The numerical results presented in this paper have been obtained
for $D_x=D_y=D_z=0.1$K.
%MIYA
%HANS
In Ref.~\cite{Irinel4} the DMI vector is taken parallel
to the $y$-axis at all the bonds and the field is applied along to the
$z$-axis. This case corresponds to the case with only $D_z$ in the
present model with the field applied in the $x$-direction.
In this case the gap opens symmetrically with field~\cite{Irinel4}.
However, as we show in this paper, the structure of the gap depends on the
direction of the field. 

\begin{figure}[t]
\begin{center}
\setlength{\unitlength}{1cm}
\begin{picture}(14,6)
\put(-1.5,0.){\includegraphics[width=8cm]{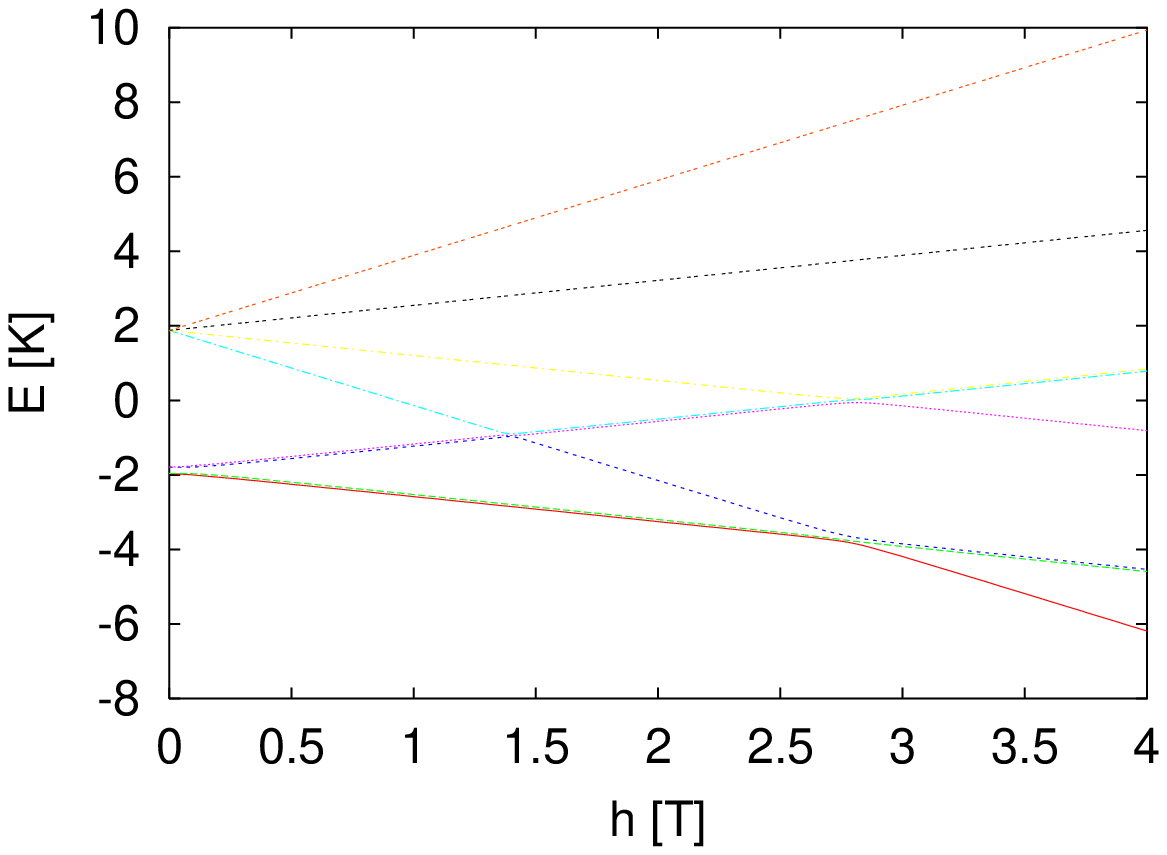}}
\put(7.,0.){\includegraphics[width=8cm]{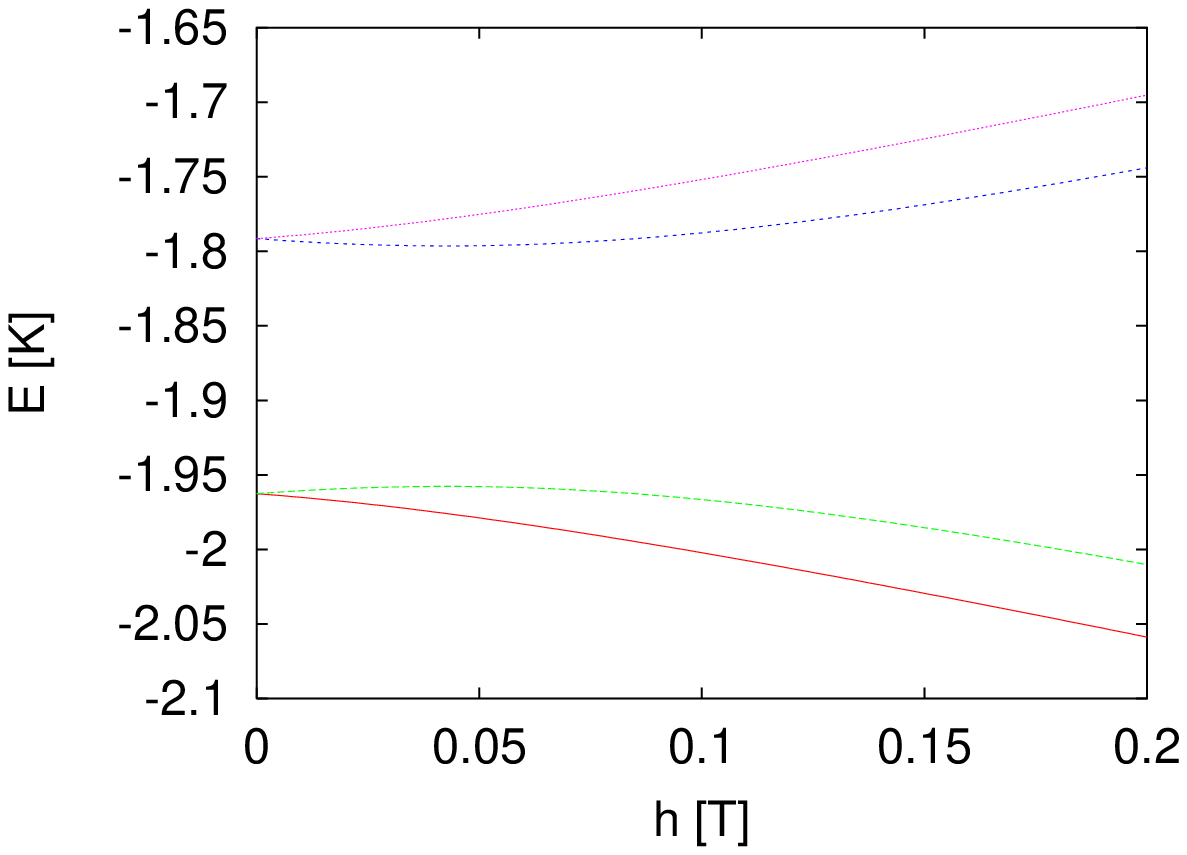}}
\end{picture}
\caption{%
Left: The eight lowest energy levels of \V\ model (\ref{VHam})
as a function of the applied magnetic field ${\bf h}$
parallel to the $z$-axis.
Right: Detailed view of the four lowest energy levels at $h\approx0$.
Note that the  energy-level splitting between the second and third
level reaches a minimum at $h\approx0.05$T, not at $h=0$.}
\label{fig11}
\end{center}
\end{figure}

In Fig.~\ref{fig11} we present results for the eight lowest energy levels of
the three-spin model (\ref{H3}) as a function of the applied magnetic field along the $z$-axis.
Qualitatively it agrees with the level diagram of the full \V\ model with parameters {\bf VsetA}.
The effect of the DMI is two-fold: as expected it lifts degeneracies but
it may also shift the position of the resonant points in a non-trivial manner.
A similar effect was also found in the full model calculations (see Sec.~\ref{sec4}).

The butterfly hysteresis loop observed in time-resolved
magnetization measurements has been interpreted in terms of combination of a LZS transition
at zero field and spin-phonon coupling~\cite{Irinel1,Irinel4}.
%MIYA
%However, a LZS transition at zero field is incompatible 
%with the idea that the DMI is responsible for splitting the S=1/2 
%quartet into two doubly degenerate levels at zero field.
%HANS: OK
Here it should be noted that unless the field is applied in a special
direction ($x$ or $y$ direction in this case), the set of avoided level crossings
is no longer symmetric with respect to the field.
Indeed, a closer look at the level diagram 
(see left picture in Fig.~\ref{fig11}) reveals that the
mimimum energy difference between the two pairs of levels does not occur at
zero field but at $h\approx0.05$T.
This implies that the LZS transition
from $|1/2,-1/2\rangle$ to the $|1/2,1/2\rangle$ level does not
take place at $h=0$ but at $h\approx0.05$T.
The minimum energy splitting between the first and second level
(counting states starting from the ground state)
also depends on the direction of the field.
For the model parameters used in our calculations, it increases from $0.05$T
for ${\bf h}$ parallel to the $z$-axis to $0.12$T
for ${\bf h}$ parallel to the $x$-axis (results not shown).
%MIYA %HANS OK
%Except for $D_z=0$ and only if ${\bf h}$ is parallel to the $z$-axis,
%there is no shift of the resonant point but then, the two lowest levels
%also remain degenerate for $h\approx0$.
%If ${\bf h}$ is no longer parallel to the $z$-axis, this degeneracy is lifted
%and the resonant point is shifted from $h=0$ to a non-zero value.
The fact that the DMI not only lifts the denegeneracy but,
depending on the direction of the field with respect to the symmetry axis,
also shifts the resonant point away from $h=0$ seems to be a generic feature.

Summarizing: Our numerical data for the parameters {\bf VsetA}, {\bf VsetB}, and {\bf VsetC}
suggest that the three-spin model reproduces the main features of the full \V\ model.
The presence of the DMI allows for adiabatic changes of the magnetization but,
according to our calculations, the value of the resonant field
for the $|1/2,-1/2\rangle$ to $|1/2,1/2\rangle$ transition
changes with the direction of the magnetic field.
%Only for very special choices of the DMI this transition takes place at $h=0$.
This change (by a factor of two at least) should lead to
observable changes in the hysteresis loops but has not been seen 
in experiment~\cite{private}.
%MIYA
%Therefore the hypothesis that the DMI is the main mechanism for generating 
%energy-level splittings in the \V\ molecule is unlikely to be correct.
%HANS : changed the English, hopefully no the content
Therefore, although the DMI causes the avoided level crossing structure,
it is anisotropic with respect to the direction of the field.
Within the three spin model we have studied the effects of higher-order correction terms
that restore the SU(2) symmetry~\cite{Kaplan,Shekntman1,Shekntman2,Zheludev}.
and found that it has no essential effect on the low energy degenerate doublets while it causes
the four $S=3/2$ levels to be degenerate at $h=0$.
In experiments only weak directional dependence was found.
Thus, another type of mechanism for the gap such as hyper-fine interaction, etc.,
is necessary and will be studied in the future.
%
% I will try more to make the paper sound `positive'.
% 

%\section{Summary}
%
\begin{acknowledgments}
We thank %B. Barbara,
I. Chiorescu, and V. Dobrovitski for illuminating discussions.
Support from the Dutch ``Stichting Nationale Computer Faciliteiten (NCF)'' is gratefully acknowledged.
\end{acknowledgments}

\section*{Appendix: Projection method}
As an alternative to the Lanczos method with full orthogonalization,
we have used a power method~\cite{WILKINSON,GOLUB} based
on the matrix exponential $e^{-tH}$~\cite{DeRaedt87}.
Writing the random vector $\Psi(0)$ in terms of the (unknown) eigenvectors
$\{\phi_i\}$ of $H$, we find

\begin{eqnarray}
\Psi(t)=e^{-tE_0} \left[
\phi_0\langle\phi_0|\Psi(0)\rangle
+e^{-t(E_1-E_0)}
\phi_1\langle\phi_1|\Psi(0)\rangle
+e^{-t(E_2-E_0)}
\phi_2\langle\phi_2|\Psi(0)\rangle
+\ldots\right],
\label{psit}
\end{eqnarray}
showing $\lim_{t\rightarrow\infty}\Psi(t)/\Vert \Psi(t)\Vert\propto \phi_0$ if
$\langle\phi_0|\Psi(0)\rangle\not=0$.
In this naive matrix-exponential version of the power method, convergence to the lowest eigenstate
is exponential in $t$ if $E_1>E_0$.

The case of degenerate ($E_0=E_1=...$) or very close ($E_0\approx E_1\approx...$) eigenvalues can be
solved rather easily by applying the projector to a subspace instead of a single vector,
in combination with diagonalization of $e^{tH}$ within this subspace~\cite{DeRaedt87}.
First we fix the dimension $k$ of the subspace by taking $k$ equal or larger than the desired number of distinct eigenvalues.
The projection parameter $t$ should be as large as possible but nevertheless
sufficiently small so that at least the first $k$ terms survive one projection step.
Then we generate a set of random initial vectors $\Psi_i(0)$ for $i=1,\ldots,k$
and set the projection count $n$ to zero.
We compute the $k$ lowest eigenstates by the following algorithm~\cite{DeRaedt87}

\begin{itemize}
\item Perform a projection step $\Psi_i((n+1)t)=e^{-tH}\Psi_i(nt)$ for $i=1,\ldots,k$.
\item Compute the $k\times k$ matrices.
$A=\langle\Psi_i((n+1)t)|e^{tH}|\Psi_i((n+1)t)\rangle=
\langle\Psi_i((n+1)t)|\Psi_i(nt)\rangle$ and
$B=\langle\Psi_i((n+1)t)|\Psi_i(n+1)t)\rangle$. Note that $A$ is hermitian and $B$ is positive definite.
\item Determine the unitary transformation $U$ that
solves the $k\times k$ generalized eigenvalue problem $Ax=\lambda Bx$. Recall that $k$ is small.
\item Compute $\Psi_i'((n+1)t)=\sum_{j=1}^k U_{i,j}\Psi_j((n+1)t)$ for $i=1,\ldots,k$.
\item Set $\Psi_i((n+1)t)=\Psi_i'((n+1)t)$ for $i=1,\ldots,k$.
\item Compute $\mu_i=\langle\Psi_i((n+1)t)|H|\Psi_i((n+1)t)\rangle$ and check if
$\Delta^2_i=\langle\Psi_i((n+1)t)|(H-\mu_i)^2|\Psi_i((n+1)t)\rangle$ is
smaller than a specified threshold for $i=1,\ldots,k$. If yes, terminate the calculation. If no, increase
$n$ by one and repeat the procedure.
\end{itemize}

We calculate $e^{-tH}\Psi$ by using the Chebyshev polynomial expansion
method~\cite{TAL-EZER,Iitaka01,LEFOR,SILVER,SlavaCheb}.
First we compute an upperbound $R$ of the spectral radius of $H$ (i.e., $\Vert H \Vert\le R$) by
repeatedly using the triangle inequality~\cite{SlavaCheb}.
From this point on we use the ``normalized'' matrix $\tilde H = (2H/R-1)/2$.
The eigenvalues of the hermitian matrix $\tilde H$ are real and lie in the interval $[-1,1]$~\cite{WILKINSON,GOLUB}.
Expanding the initial value $\Psi(0)$ in the (unknown) eigenvectors $\phi_j$ of $\tilde H$ (or $H$)
we find
\begin{equation}
\Psi(t)=e^{-tH}\Psi(0)=
e^{z\tilde H}\Psi(0) =\sum_j e^{z\tilde E_j} \phi_j
\langle\phi_j|\Psi(0)\rangle,
\label{expz}
\end{equation}
where $z=-tR$.
We find the Chebyshev polynomial expansion of $\Psi(t)$ by computing the
Fourier coefficients of the function $e^{z\cos\theta}$~\cite{ABRAMOWITZ}.
Alternatively, since $-1\le \tilde E_j \le 1$, we can use the expansion
$
e^{z\tilde E_j}=I_0(z) + 2\sum_{m=1}^{\infty} I_{m}(z)T_{m}(\tilde E_j)
$
where $I_m(z)$ is the modified Bessel function of integer order $m$~\cite{ABRAMOWITZ}
to write Eq.~(\ref{expz}) as
\begin{equation}
\Psi(t)
=\left[I_0(z)I + 2\sum_{m=1}^{\infty} I_{m}(z) T_{m}(\tilde H)\right] \Psi(0)\,.
\label{SUM0}
\end{equation}
Here, $I$ is the identity matrix and $T_{m}(\tilde H)$ is the
matrix-valued Chebyshev polynomial defined by the recursion
relations
\begin{equation}
 T_{0}(\tilde H)\Psi(0)=\Psi(0)\label{CHEB1}\,,\quad
 T_{1}(\tilde H)\Psi(0)=\tilde H\Psi(0)\,,
\label{CHEB44}
\end{equation}
and
\begin{equation}
 T_{m+1}(\tilde H)\Psi(0)=
2\tilde H T_{m}(\tilde H)\Psi(0)- T_{m-1}(\tilde H)\Psi(0)\,,
\label{CHEB4}
\end{equation}
for $m\ge1$.
In practice we will sum only contributions with $m\leq M$ where $M$ is choosen such that
for all $m>M$, $|I_m(z)/I_0(z)|$ is zero to machine precision.
Then it is not difficult to show that
$\Vert e^{-tH}/I_0(z) - I - 2\sum_{m=1}^{M} [I_{m}(z)/I_0(z)] T_{m}(\tilde H)\Vert$
is zero to machine precision too (instead of $e^{-tH}$ we can equally well
use $e^{-tH}/I_0(z)$ as the projector).

Using the downward recursion relation of the modified Bessel functions,
we can compute $K$ Bessel functions to machine precision using only
of the order of $K$ arithmetic operations~\cite{ABRAMOWITZ,NumericalRecipes}.
A calculation of the first 20000 modified Bessel functions takes less than 1 second
on a Pentium III 600 MHz mobile processor, using 14-15 digit arithmetic.
Hence this part of a calculation is a negligible fraction of the total computational work
for solving the eigenvalue problem.
Performing one projection step with $e^{-tH}$ amounts to repeatedly using recursion (\ref{CHEB4}) to
obtain $\widetilde T_{m}(B)\Psi(0)$ for $k=2,\ldots,M$, multiply the elements
of this vector by $I_{m}(z)$ and add all contributions.

\newpage

\end{document}